# Passive wing deployment and retraction in beetles and flapping microrobots

Hoang-Vu Phan[1,✉], Hoon Cheol Park[2,3], and Dario Floreano[1,3]

[1]School of Engineering, Ecole Polytechnique Fédérale de Lausanne (EPFL), CH-1015 Lausanne, Switzerland

[2]Department of Smart Vehicle Engineering, Konkuk University, Seoul 05029, South Korea

[3]These authors contributed equally

✉Corresponding email: vu.phan@epfl.ch; hvphan11@gmail.com

## Abstract

Birds, bats and many insects can tuck their wings against their bodies at rest and deploy them to power flight. Whereas birds and bats use well-developed pectoral and wing muscles and tendons[1,2], how insects control these movements remains unclear, as mechanisms of wing deployment and retraction vary among insect species. Beetles (Coleoptera) display one of the most complex wing mechanisms. For example, in rhinoceros beetles, the wing deployment initiates by fully opening the elytra and partially releasing the hindwings from the abdomen. Subsequently, the beetle starts flapping, elevates the hindwings at the bases, and unfolds the wingtips in an origami-like fashion. Whilst the origami-like fold have been extensively explored[3–7], limited attention has been given to the hindwing base deployment and retraction, which are believed to be driven by thoracic muscles[4,8–10]. Using high-speed cameras and robotic flapping-wing models, here we demonstrate that rhinoceros beetles can effortlessly elevate the hindwings to flight position without the need for muscular activity. We show that opening the elytra triggers a spring-like partial release of the hindwings from the body, allowing the clearance needed for subsequent flapping motion that brings the hindwings into flight position. The results also show that after flight, beetles can leverage the elytra to push the hindwings back into the resting position, further strengthening the hypothesis of a passive deployment mechanism. Finally, we validate the hypothesis with a flapping microrobot that passively deploys its wings for stable controlled flight and retracts them neatly upon landing, which offers a simple yet effective approach to the design of insect-like flying micromachines.

## One sentence summary

Beetles can effortlessly deploy and retract the hindwings without the need for thoracic muscles, inspiring the design of a compact flapping microrobot with actuator-free, self-deploying, self-retracting wings.

**Main text**

The wings of flying insects are vulnerable and fragile structures that are crucial for evading predators, foraging, migrating or mating. Most insects are therefore capable of folding and resting their wings against the sides of the abdomen to reduce wing damage risks and interference in terrestrial locomotion through narrow spaces. Mechanisms of wing deployment and retraction vary among insect species due to absence of intrinsic wing muscles and to differences in wing shape, structure, and function[11,12]. In particular, beetles (Coleoptera) display one of the most complex mechanisms among various insect species[4,7]. Beetles have two distinct pairs of wings: a pair of membranous and fragile wings (hindwings) and a pair of hardened forewings (elytra) used mostly to protect the hindwings at rest (Fig. 1a,b). Hindwings are an origami-like foldable structure that allows them to neatly stow between the body and the elytra and deploy to power flapping flight. The complete hindwing deployment procedure consists of elevating the wing base and unfolding the wingtip. Studies of the hindwing have primarily focused on origami-like fold of the wingtip and proposed the use of elastic elements[3,13], thoracic muscles[4,5], hydraulic mechanism[6], or flapping forces[7] to drive the unfolding. However, the mechanism leveraged by beetles to elevate the hindwing bases to flight position and bring them back to rest against the body remains little understood. The most common explanation is that beetles, as well as other insects in the group of Neoptera such as wasps, bees, and flies, use direct flight muscles attached to the basilar sclerite and the third axillary sclerite of the wing base to drive these movements[4,8–10]. However, there is no experimental evidence showing muscle activity during hindwing deployment and retraction.

**Muscle-free passive hindwing deployment**

To gain insight into how a beetle elevates its hindwings at the bases, we used synchronized high-speed cameras to record wing deployment kinematics of rhinoceros beetles *Allomyrina dichotoma* (Fig. 1c–g and 'Wing kinematics experiments' in Methods). We observed that the beetle initiates a flapping flight with a two-phase wing deployment. In the first phase, the beetle fully elevates the elytra, followed by a partial release of the hindwings to an angle of about 48.5 ± 0.7º (n = 7) from the abdomen while maintaining the wingtip in the folded configuration (Fig. 1c and Supplementary Video 1). As the tips of the left and right hindwings overlap when stowed against the abdomen (Fig. 1a), the right hindwing may not be released instantly upon the opening of the right elytron if the left elytron and hindwing remain folded, consequently affecting the right hindwing's release time and speed (Fig. 1d,e and Extended Data Fig. 1). After release, the hindwing undergoes a series of decreasing amplitude oscillations until it comes to rest at an equilibrium position, showing characteristic of an underdamped spring-mass system (Fig. 1d). The amplitude of the oscillation is proportional to the speed of the initial release (Fig. 1e, and Extended Data Fig. 1). We thus suggest that the first phase of the hindwing elevation is passively triggered by the release of stored elastic energy rather than by active muscular control.

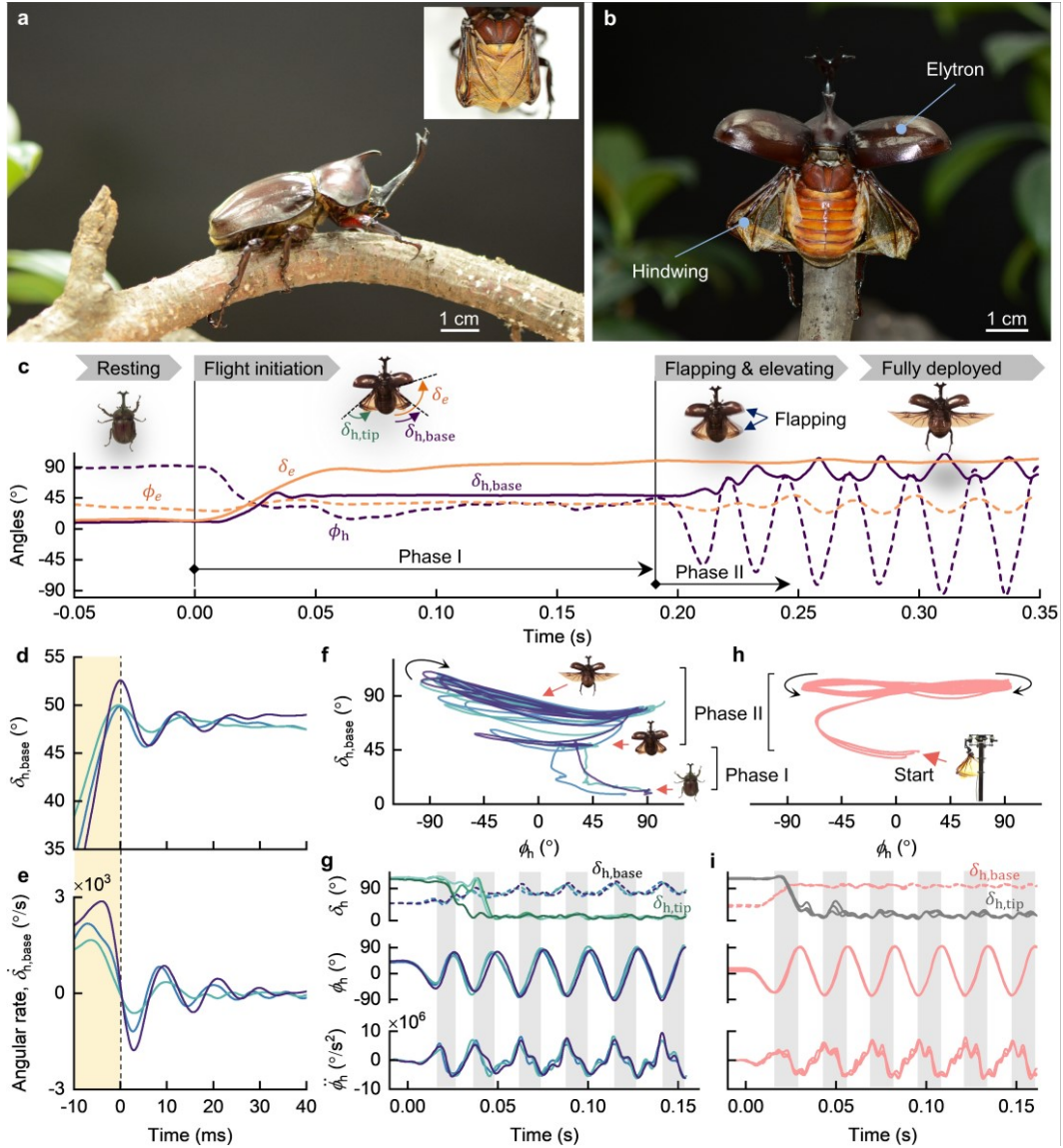

**Fig. 1. Muscle-free, passive hindwing deployment in the beetles *Allomyrina dichotoma*. a,** During ground locomotion or at rest, beetles stow neatly their hindwings under the elytra**.** The inset shows the tips of the left and right hindwings stacked on top of each other when in folded configuration. **b,** The beetle prepares for flight by fully opening the elytra and partially releasing the hindwings from the abdomen. **c,** Deployment kinematics of the elytron and hindwing from rest to flight position. Orange, elytron; violet, hindwing; solid line, elevation angle at the wing base ($\delta$); dashed line, stroke angle ($\phi$). **d,e** Partial elevation angle (**d**) and angular velocity (**e**) of the hindwing after the elytron released, showing characteristic behavior akin to an underdamped spring-mass system. Shaded area denotes variable-velocity elevation of the hindwing. **f,g,** Hindwing flapping and deployment kinematics powered by a beetle (n = 3) in a form of wing leading edge trajectory with the shape of a half-fold figure-eight (**f**) and as a function of time (**g**). **h,i,** Hindwing flapping and deployment kinematics (n = 3) driven by a flapping mechanism in a form of wing leading edge trajectory with the shape of an inverted figure-eight (**h**) and as a function of time (**i**). Gray areas in **g** and **i** denote the upstroke flapping motions. The time instant is set to 0 s when the elytron starts elevation in **c**, the elevation angle of hindwing reaches the peak in its first release in **d**, and the hindwing starts flapping in the second deployment phase in **g** and **i**.

The beetle then initiates the second wing deployment phase by activating synchronized flaps of both wing pairs, followed by an outward elevation of the hindwing bases and an unfolding of the origami-like wingtips (Fig. 1c,f,g) that bring the hindwings into flight position. We observed that this deployment sequence is initiated only when the wing starts to flap, which is activated without any correlation with the timing of initial hindwing elevation in the first phase (Extended Data Fig. 1d). We hypothesize that the beetle leverages flapping forces not only to unfold the hindwing tip as previously shown[7], but also to passively elevate the hindwing at the base. To test this hypothesis, we affixed a freshly removed hindwing from the beetle onto a flapping mechanism with active flapping and an actuation-free elevation joint at the wing base ('Wing kinematics experiments' in Methods and Extended Data Fig. 2). We then activated the hindwing to flap at a flapping frequency similar to that of the beetle. We found that the hindwing passively elevates at the base within the first flapping cycle (Fig. 1h,i, and Supplementary Video 2), even when experiencing different flapping frequencies (Extended Data Fig. 2). During the first flapping cycle when the wingtip remains folded, hindwing elevation is driven mostly by the centrifugal force[14], $F_c = m_w \dot{\phi}^2 r_{w,CG}$, in which $m_w$ is the wing mass and $r_{w,CG}$ is the distance between the flapping axis and the center of mass of the wing. The contributions of other force components, i.e. wing inertial force and aerodynamic drag, which are tangential to the flapping motion, are negligible because they are orthogonal to the elevation motion plane. Since $F_c \propto r_{w,CG}$, the first phase of the hindwing release (phase I in Fig. 1c) enables unobstructed flapping motion and facilitates the centrifugal acceleration to induce the subsequent elevation phase (phase II in Fig. 1c). Although the trajectory of the hindwing driven by the robotic mechanism did not completely match that of beetles, these results support the hypothesis that beetle can effortlessly elevate the hindwing without muscular activity (Fig. 1h,i).

**Elytron-driven hindwing retraction**

After flight, beetles retract elytra and hindwings at the wing bases against the abdomen, subsequently folding the hindwing tips to tuck neatly under the elytra at rest (Extended Data Fig. 3). Beetles can leverage the abdomen and elytra to pull the tips of the hindwings into fully folded configuration[3,7,13]. Whereas active thoracic muscles have been proposed to drive the retraction of the hindwing at the base[5,9,10]. However, if rhinoceros beetles do not need thoracic muscles to deploy their hindwings, the question arises if they can also retract the hindwings without those muscles. To address this question, we analyzed wing kinematics of tethered beetles at the end of flapping flights (Fig. 2, and 'Wing kinematics experiments' in Methods). To terminate the flight, the beetle gradually decreases the flapping frequency and the peak-to-peak stroke amplitude of the hindwings, resulting in a decrease in the elevation angle (Fig. 2a,b, and Extended Data Fig. 3). Thereafter, the beetle initiates the retraction of its elytron but keeps the hindwing elevated until the foremost edge of the elytron and the leading edge of the hindwing make contact (t = 0 s in Fig. 2b,c, and Supplementary Video 3). From the time of contact, the elytron and the hindwing come together towards the abdomen and fully close in within 70 ms (Fig. 2c (left wings) and Fig. 2d,f). From these observations, we hypothesize that the beetle leverages its elytra to depress the hindwings down rather than using active muscular control at the hindwing's bases. To test this hypothesis, we investigated the retraction kinematics of the hindwing in the absence of the elytron ('Wing kinematics experiments' in Methods). We observed that, without the elytron, the hindwing remains elevated and cannot retract (Fig. 2c (right wing), Fig. 2e,g, and Supplementary Video 4). Furthermore, in this condition, the beetle attempted to use its legs as an alternative method to depress the hindwing to the resting position (Supplementary Video 5). These findings thus suggest that elytra not only offer previously known protective capabilities[15], and enhanced aerodynamic lift[16,17] and flight stability[18,19], but also serve to retract the hindwings after flight. These results show that thoracic muscles are not needed to fold the hindwing base against the abdomen and further support the hypothesis that they may not be used for deployment of the hindwing.

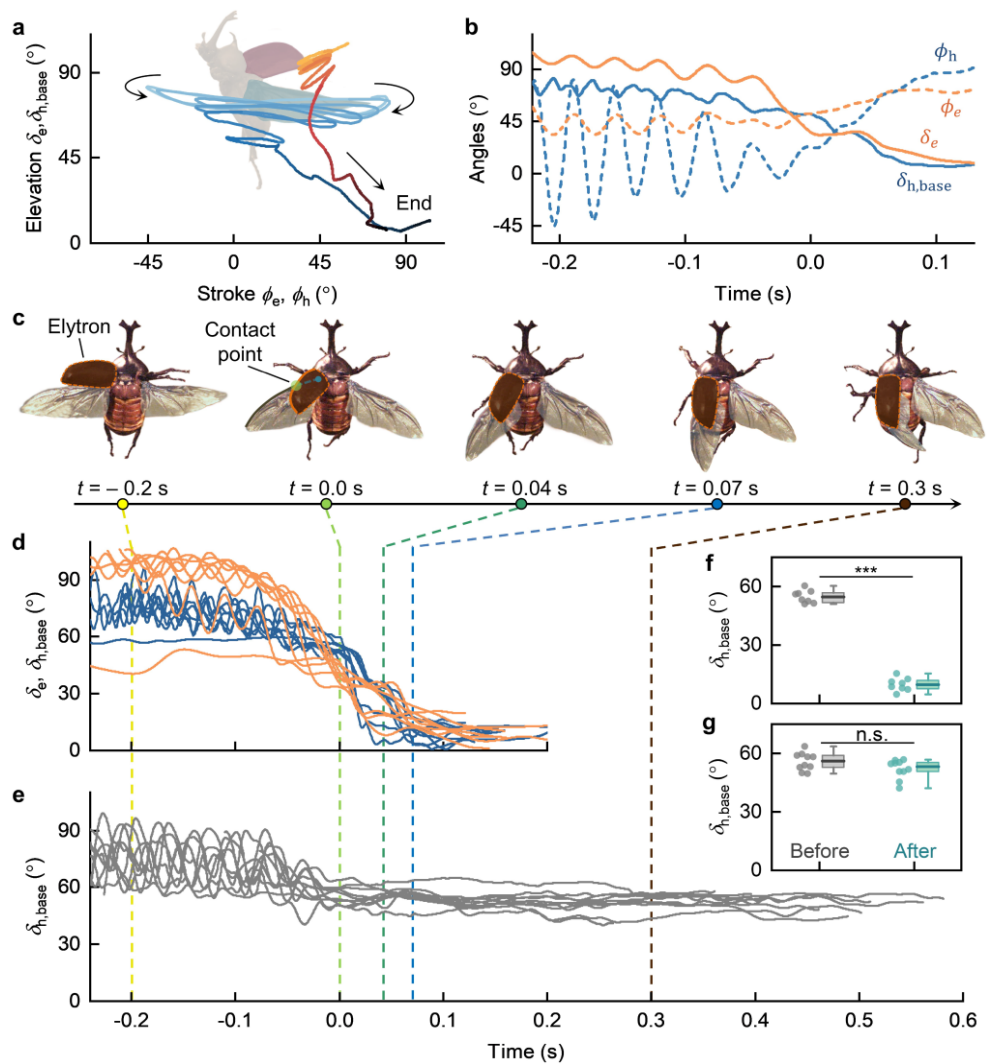

**Fig. 2. Beetles can leverage their elytra to push the hindwings back to the abdomen after flight without the need for the hindwing's internal muscular activity. a,b**, Retracting kinematics of the elytron (orange) and hindwing (blue) in the form of wing trajectory (**a**) and as a function of time (**b**). The hindwing trajectory shows the shape of an inverted figure-eight, similar to the hindwing deployment trajectory driven by the flapping mechanism in Fig. 1h. In **b**: solid line, elevation angle at the wing base ($\delta$); dashed line, stroke angle ($\phi$). **c**, Sequential sketches illustrate the closing trajectories of the hindwing with presence (left side) and absence (right side) of the elytron. **d**, During closing, the elytron (orange) can push the hindwing (blue) back to the abdomen (n = 8). **e**, In absence of the elytron, the hindwing is unable to fold back after flight (n = 10). **f,g**, Comparison of mean elevation angles of the hindwing before (black) and after (cyan) closing with presence (**f**) and absence (**g**)

of the elytron (median, interquartiles, and range). ***$P < 0.001$; n.s., not significant (two-sample, two-tailed *t* test). The time instant is set to 0 s when the elytron touches the hindwing while closing.

**Self-deployable, self-retractable robotic wing**

We validated the passive deployment and retraction mechanism with a flapping microrobot. Over the past decade, numerous flapping-wing robots that mimic insects at various scales have been developed[20–28], but they all use flapping wings locked in fully extended configuration that cannot be retracted after flight as insects do. Although beetle-inspired folding wings have been used in propeller-driven drones[29,30], they used servomotors for active deployment and folding of the wings. Here we designed a novel mechanism for flapping wings that can passively retract and deploy by flapping motion (Fig. 3, and 'Deployable and retractable wing design' in Methods) as suggested by the aforementioned experimental analyses. The wing consists of a leading-edge spar, a wing membrane, and a hinge joint at the wing base (Fig. 3a–c). To replace the role of the beetle's elytron, which would add complexity and mass to the robot, the wing is equipped with an elastic tendon at the armpit, which facilitates rapid wing closing in within 100 ms but still allows wing release in one flapping cycle at various frequencies (Fig. 3c–e, and Supplementary Video 6).

We first tested the effect of the centrifugal force induced by the wing inertia on the wing elevation by replacing the wing membrane with a proportionally distributed mass along the wing's leading-edge spar. In the presence of the armpit tendon, the wing is required to flap at a sufficiently high frequency (> 14 Hz) to overcome the elastic force of the tendon and remain on the plane perpendicular to the flapping axis ($\delta = 90°$), despite a small downward deviation at the beginning of each stroke (Fig. 3e, and Extended Data Fig. 4). These results indicate that the centrifugal force induced by the wing inertia during flapping is sufficient to elevate the wing, consistently with previous findings[14,31]. However, when adding the membrane, the wing cannot maintain the expected elevation angle of 90° during flapping motion; it deviates downward from the threshold ($\delta_{threshold} = 90°$) with a maximum deviation angle of $|\Delta\delta_{max}| = 12.1°$ (Fig. 3f, and Supplementary Video 7). As a result, the robot generated lower vertical forces than in the condition with non-retractable wings (Fig. 3h). To mitigate the downward deviation, we increased the threshold of the elevation angle to 100° to allow the wing's center of gravity, located at approximately 20% of the wing chord from the leading edge, to stay on the plane perpendicular to the flapping axis during flapping (Fig. 3f,g). This increment thus maximizes the effect of the centrifugal acceleration that helps elevate the wing by increasing the distance from the flapping axis to the wing's center of gravity. We found that the downward deviation of the wing is reduced to $|\Delta\delta_{max}| = 6.3°$, which is comparable to the maximum elastic deflection of the leading-edge bar in the non-retractable wing ($|\Delta\delta_{max}| = 4.5°$) (Fig. 3g, and Extended Data Fig. 4). Moreover, it enables the robot to generate vertical forces similar to those generated by the non-retractable wing (Fig. 3h), helping to preserve flight capability with all onboard components.

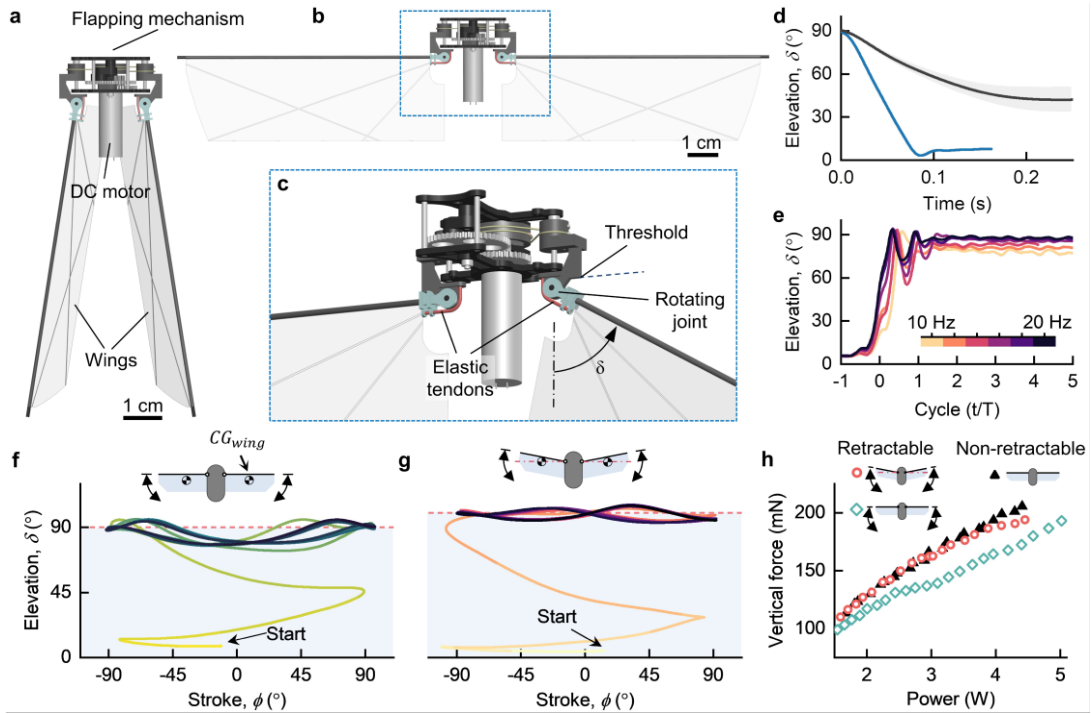

**Fig. 3. Insect-inspired flapping wings that deploy for powered flight and fold back to rest without the need for additional actuators**. **a,b,** Flapping wings in folded (**a**) and extended (**b**) configurations. **c**, An elastic tendon (shown in red) at the armpit enables passive folding of the deployable wing. Maximum elevation angle is limited by the threshold. **d**, Elastic tendons allow the wing to fold back rapidly within 100 ms. Blue, with elastic tendon; black, no elastic tendon. **e**, Wing elevation driven by only wing inertia at different flapping frequencies with presence of the elastic tendon at the armpit. **f,g,** Wing leading-edge trajectory in terms of the elevation and stroke angles for the retractable wings with the thresholds of $\delta_{threshold} = 90°$ (**f**) and $\delta_{threshold} = 100°$ (**g**). The reaction force induced by the wing colliding with the threshold at the end of the previous stroke may contribute to the downward movement of the wing at the beginning of the subsequent stroke in **f**. The flapping frequency was 20 Hz. **h**, Cycle-averaged vertical force as a function of input power for three wing configurations.

**Untethered flight demonstrations**

We integrated the passively deployable and retractable wings into an insect-inspired, tailless, flapping-wing microrobot[7,26], which can decrease the tip-to-tip wingspan from 20 cm to only 3 cm when fully folded (Fig. 4a). Upon activation of the flapping motor, the wings elevated to the flapping plane within two flapping cycles and retracted to rest within 100 ms of motor deactivation (Fig. 4b, and Supplementary Video 8). To test whether the flapping wings enable stable flight, we conducted experiments with untethered takeoff, hovering, and landing (Fig. 4c–i, 'Flight experiments' in Methods, Extended Data Figs. 5 and 6, and Supplementary Video 9). The results showed that when the flapping motion was activated, the wings passively elevated from the fully folded configuration to the flight position and generated sufficient lift for takeoff (Fig. 4c). Despite experiencing slight oscillations in roll (root-mean-square error, RMSE = 7.8°) and pitch (RMSE = 4.7°) angles and a drift in yaw angle (heading), the robot could successfully hover and maintain a stable upright configuration while airborne (Fig. 4d,f–i, and Extended Data Fig. 6). When deactivating the flapping motion upon landing, the wings passively and rapidly folded back against the robot's body (Fig. 4e). We also show that if the wings hit an obstacle in flight, which causes the robot to destabilize and tumble, they rapidly retract against the body before reaching the ground, thus helping to prevent wing damage (Fig. 4j, and Supplementary Video 10). In summary, these experiments not only validate the hypothesis of passive deployment and retraction of beetle wings, but also demonstrate its translation into a new design principle for robust flight of flapping-wing microrobots with stringent weight constraints in cluttered and confined spaces.

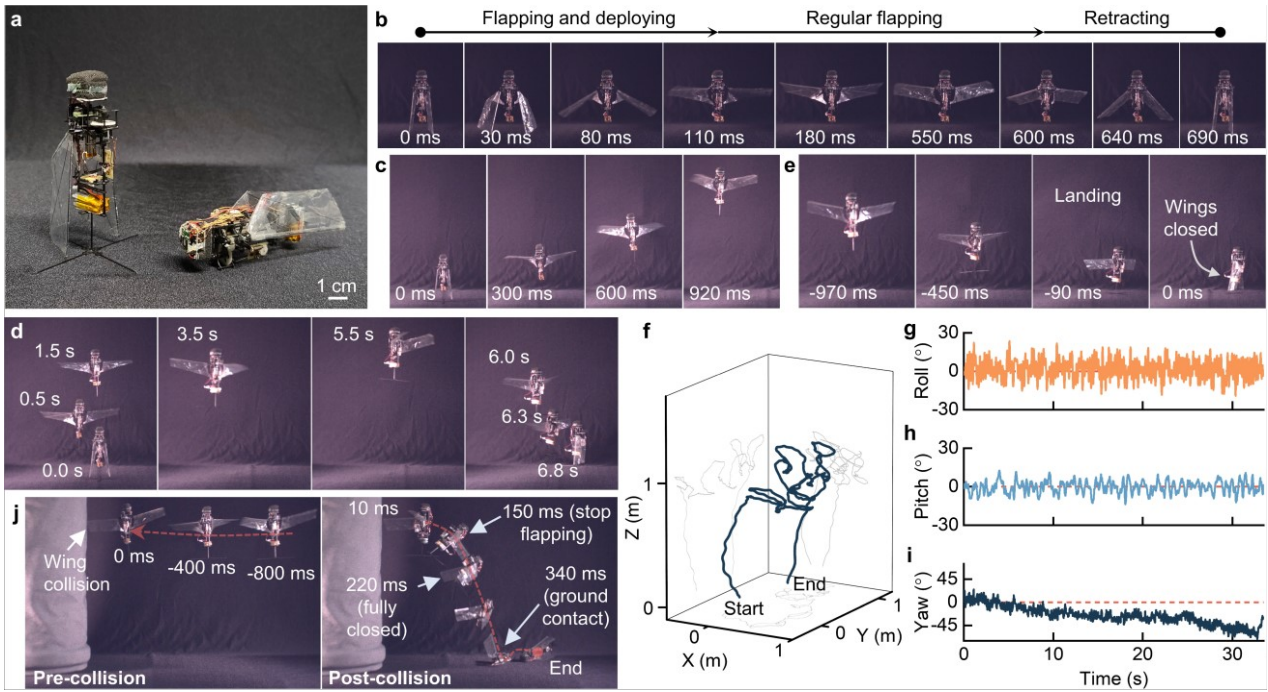

Fig. 4. Insect-inspired flapping microrobots can sustain untethered controlled flight with self-deployable and self-retractable wings. **a,** The 18-gram tailless flapping robots with passive deploying-retracting wings. **b,** The robot can passively deploy and retract its wings through activation and deactivation of the flapping motion, respectively. **c–e,** Composite images of the robot during takeoff (**c**), hovering flight (**d**), and landing (**e**). **f–i,** Three-dimensional flight trajectory (**f**), and body attitude angles: roll (**g**), pitch (**h**), and yaw (**i**). The red dashed lines in **g–i** denote the reference of 0°. The robot experienced a drift in yaw angles as it was stabilized by angular rate signal only. **j,** Rapid retraction protects the wings from a crash-landing due to inflight wing collisions. The time instant is set to 0 ms when the wing collides with the wall.

## Conclusions

In summary, our results reveal that beetles can leverage elastic energy and flapping forces to passively deploy hindwings for flight and the elytra to push them back to rest, rather than relying on a distinct group of thoracic muscles. Although beetles can deploy and close their elytra by muscular control[32], the different deployment patterns and the passive activation of the hindwings suggest that the deployment-and-retraction mechanisms in the two wing pairs are uncoupled. Our findings suggest that beetles recruit pre-existing mechanisms (flapping and elytron) to decrease muscular activities, and open the door for additional studies to investigate to what extent other small-scale flying insects may leverage similar strategies. We also translated the principle into a passive deployment and retraction mechanism for flapping-wing microrobots and showed that the robot can passively deploy wings for take-off, perform stable hovering, and rapidly retract the wings against the body upon landing or in case of in-flight collisions without the need for additional actuators. The findings and results thus help advance our understanding of effective locomotion strategies in insects and have implications for flapping-wing robots, particularly those at micro-scales with limited takeoff weights[33,34].

## Methods

**Insects**

Ten adult male and female Rhinoceros beetles, *Alloymyrina Dichotoma*, with body masses of 6–9 g were purchased from a local store in Korea in July 2022. We then reared the beetles in a plastic cage (0.5 m × 0.4 m × 0.4 m) at room temperature (25°C) with jelly foods and flake soil. All experiments on beetles were conducted at Konkuk University following the institutional and national guidelines for the use of laboratory animals. Only beetles capable of free flight were used for the tethered flight experiments.

**Wing kinematics experiments**

We used three calibrated high-speed cameras (Photron Ultima APX, frame rate of 2,000 fps, resolution of 1,024 × 1,024 pixels, and shutter speed of 4,000 fps) to film the three-dimensional wing deployment-retraction kinematics of the tethered beetles and flapping-wing robots. More details on the experimental setup and camera calibration can be found in Phan and Park[7]. We fixed the back head of a beetle to a fixture to facilitate tethered flapping flight. To digitize the wing kinematics of the beetles, we placed markers (1 mm diameter white ink) on the elytra, the hindwings and the body. For the experiments on hindwing closure without the elytra, we removed the right elytron of the beetles at the base.

We used the open-source MATLAB-based DLTdv digitizing tool[35] to track the markers recorded by the three synchronized cameras and obtain their three-dimensional coordinates, smoothed with 95% confidence interval. Using these coordinates, we can determine the stroke angles ($\phi$, defined as the angle between the extended line connecting the bases of the left and right wings and the wing's leading edge) and the elevation angles ($\delta_{base}$, defined as the angle between the side of the beetle's abdomen (the flapping axis in the robot) and the wing's leading edge) of the elytron and the hindwing. The folding angle of the hindwing tip, $\delta_{h,tip}$, which is the angle between the radius anterior (inner segment of the hindwing's leading edge) and the radius anterior 3 (outer segment of the hindwing's leading edge) (Fig. 1a), was also calculated.

After recording the wing kinematics of the beetles, we detached the hindwing from the beetles and attached it to a motor-driven flapping mechanism by gluing the hindwing's base to the flapping crank, which allows both actuated flapping and actuation-free elevation at the wing base (Extended Data Fig. 2). To generate a symmetric downstroke and upstroke wing motion with high stroke amplitude as found in beetles[36], we built a flapping mechanism that combines the scotch-yoke and the pulley-string mechanisms (Extended Data Fig. 2). We used an external power supply (E36103A, Keysight) to activate the flapping motion of the hindwing at a flapping frequency similar to that of the beetle. We then conducted the wing kinematics experiments with the same procedure used for the beetles to obtain the stroke angle $\phi$, elevation angle $\delta_{base}$, and the folding angle $\delta_{h,tip}$, of the hindwing driven by the flapping mechanism.

**Deployable and retractable wing design**

The wing consists of a straight leading edge made of 1 mm carbon tube, a 10 μm Mylar membrane surface reinforced by 0.3 mm carbon rod for optimal aerodynamic performance[26], and a lightweight rotating hinge joint at the wing base made of carbon/epoxy panels with thicknesses of 0.5 mm and 1.0 mm (Fig. 3a–c). The membrane can freely rotate around the leading edge during flapping motion. One end of the hinge joint (cyan color in Fig. 3c) is fixed to the 1 mm leading edge, while the other end (black color in Fig. 3c) attached to the output linkage of the flapping mechanism serves as a threshold to limit the elevation of the wing. The wing's center of gravity is located at about 20% wing chord from the leading edge and at about mid-wingspan. To ensure that the center of gravity lies on the plane perpendicular to the flapping axis during flapping, thereby maximizing the centrifugal effect, we increased the threshold limit of the elevation angle from 90° to 100°.

Adding an elytra-like mechanism to drive the wing retraction increases mechanical complexity and mass of the robot. To enable passive closing after flapping flight, we alternatively equipped the wing with an elastic tendon at the hinge joint (Fig. 3c). To ensure that the elastic force of the tendon is high enough to retract the wing in any orientation of the robot, but still allows flapping forces to elevate the wing to the flight position rapidly, we used the tendon with an elastic constant, $k_e$, that satisfies the condition:

$$(m_w g r_{w,CG} + T_{mem}) < k_e \Delta l r_e < (F_c l_{w,CG} + F_{v,aero} r_{aero} - m_w g r_{w,CG}), \qquad (1)$$

where $m_w$, $g$ and $r_{w,CG}$ are the wing mass, gravity, and the distance from the flapping axis to the center of the wing mass, respectively, $T_{mem}$ is the torque due to the deflection force of the wing membrane, $\Delta l$ denotes the displacement of the tendon, $r_e$ is the distance from the elevation axis to the tendon, $F_c = m_w \dot{\phi}^2 r_{w,CG}$ is the centrifugal force, $l_{w,CG}$ is the distance from the center of the wing mass to the wing sweeping plane perpendicular to the flapping axis, $F_{v,aero}$ represents the vertical aerodynamic force, and $r_{aero}$ denotes the distance from the flapping axis to the center of aerodynamic force. For the case of wing inertia only, the aerodynamic term ($F_{v,aero} r_{aero}$) was excluded. We also assume that the friction at the joint is negligible. We tested various tendons (elastic strings) with different elastic constant values to find a proper one.

**Vertical force and power measurements**

We used a 6-axis Nano 17 force/torque sensor (ATI Industrial Automation, force resolution ≈ 3 mN) to measure the vertical forces generated by the flapping mechanism with three wing configurations: deployable-retractable wings with the thresholds of $\delta_{threshold} = 90°$ and $\delta_{threshold} = 100°$, and non-retractable wing. We mounted the flapping-wing mechanism on the Nano 17 sensor, where the z-axis of the mechanism is aligned with the Z-axis of the sensor. We used an external DC power supply (PeakTech 6226) to power the flapping-wing mechanism at different input voltages and thus different flapping frequencies. The force data were recorded by the ATI DAQ F/T software (ATI Industrial Automation) at the sampling frequency of 3200 Hz during a 5-s flapping period. Along with the force, we also measured input power of the flapping-wing mechanism ($P_{flap}$). We connected a resistor, $R = 1$ Ohm, in between the power supply and the flapping-wing mechanism. We then used an oscilloscope (HMO2024, Rohde & Schwarz) to measure input ($V_{R,in}$) and output ($V_{R,out}$) voltages of the resistor to obtain the current, $I = (V_{R,in} - V_{R,out})/R$, during the flapping motion. Thus, $P_{flap} = V_{flap}I$, where $V_{flap} = V_{R,out}$.

**Robot prototype**

We prototyped the 18-gram flapping-wing robot described previously[7,26] to demonstrate its untethered controlled flights with the passively deployable and retractable wings. The robot consists of a transmission system to convert the rotary motion of a DC motor (Chaoli CL720) to a high-amplitude flapping motion of the wings; a three degrees-of-freedom mechanism of attitude control and stabilization driven by three micro servomotors (LZ servo) that can modulate the flapping stroke plane for pitch and roll controls and the wing root spars for yaw control; an avionic system with an Arduino-based microcontroller (Bareduino Nano, Seeedstudio), a 6-axis IMU (MPU6050) and a receiver (Deltang DT-Rx36); and a two-cell 70 mAh lithium-polymer battery (Hyperion). As our tailless flapping-wing robot is inherently unstable after takeoff[20,21,33], we stabilize its flight by using a proportional-derivative (PD) controller[7,26] to sense the attitude angles and angular rates of the robot measured by the onboard IMU.

**Flight experiments**

Before the flight experiments, we locked the wings in a fully elevated configuration and trimmed the robot by conducting tethered takeoffs without activating feedback signals and control inputs to eliminate any initial pitch and roll moments caused by imperfect fabrication. We also conducted tethered flights of the robot and used trial-and-error method to fine-turn the control gains. We used a high-speed camera (Chronos, frame rate of 500 fps, resolution of 1,024 × 1,024 pixels) to film the wing flapping motion, and takeoff, hovering and landing flights of the flapping-wing robot in Fig.4b-e,j. To obtain the flight trajectory and body attitudes of the robot, we placed lightweight reflective markers on the robot (Extended data Fig. 5). The robot was remotely piloted in a 10 m × 10 m × 8 m indoor flight arena equipped with a motion tracking system (26 Optitrack cameras, 120 Hz) to track these markers and convert to the three-dimensional trajectory and body attitude angles of the robot following the roll–pitch–yaw rotation sequence.

# Acknowledgements

This project was partially funded by the Swiss National Science Foundation through the NCCR Robotics program, and by the Korea government (MSIT) (No.2022R1A4A101888411) through the National Research Foundation of Korea (NRF).

# Author contributions

H.-V.P. conceived the idea and designed the research, designed and built the robot, performed all experiments on insects and robots, processed and analyzed the data, and originally drafted the manuscript. H.C.P and D.F contributed to the data analysis and writing of the manuscript. All authors gave final approval for publication.

# Data availability

The data that support the findings of this study are available within the paper and its Supplementary Information. Other datasets are available from the corresponding author on reasonable request.

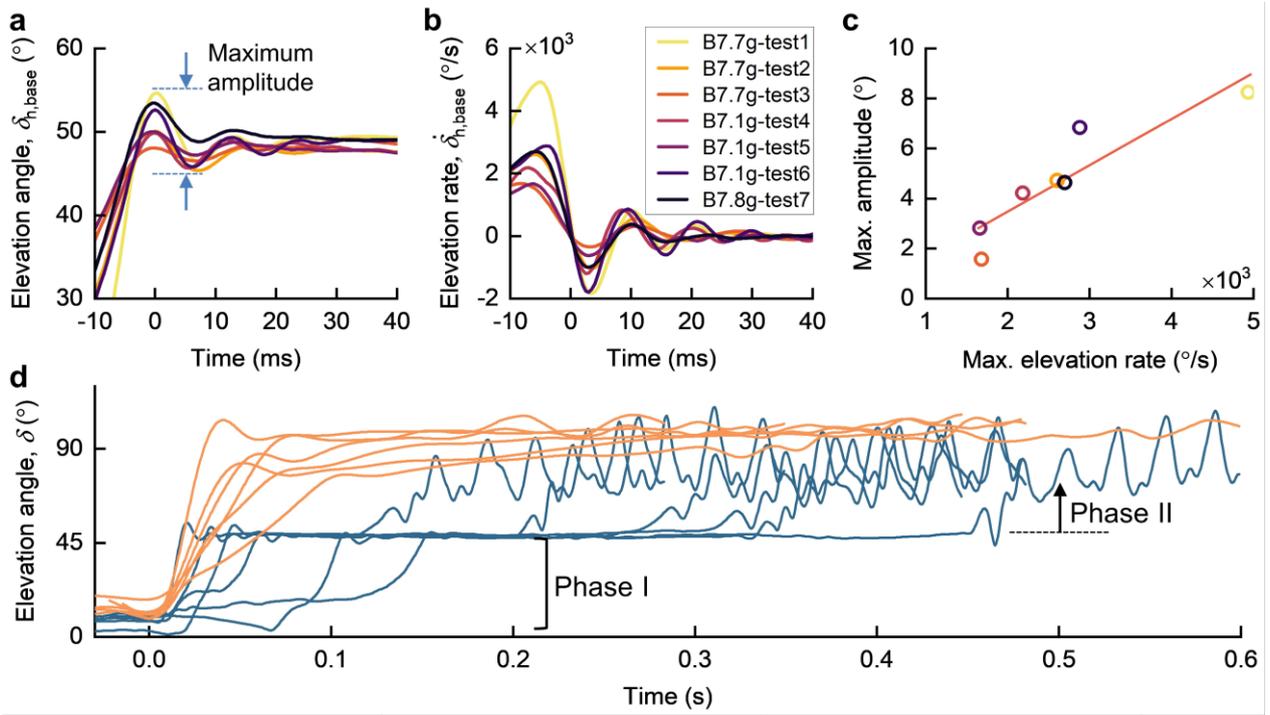

**Extended Data Fig. 1| Wing deployment kinematics of beetles. a,b**, At the end of the partial release (phase I of the deployment), the hindwing experiences decreasing oscillations around the equilibrium position, $\delta_{h,base}$ = 48.5 ± 0.7°: (**a**) elevation angle (**b**) and angular rate. **c**, Amplitude of the first oscillation (shown in **a**) is proportional to the releasing rate of the hindwing in **b**. Red line denotes the linear fit. **d**, Elevation angles of the elytron (orange) and the hindwing (dark blue). The time instant is set to 0 s when the elytron starts to elevate.

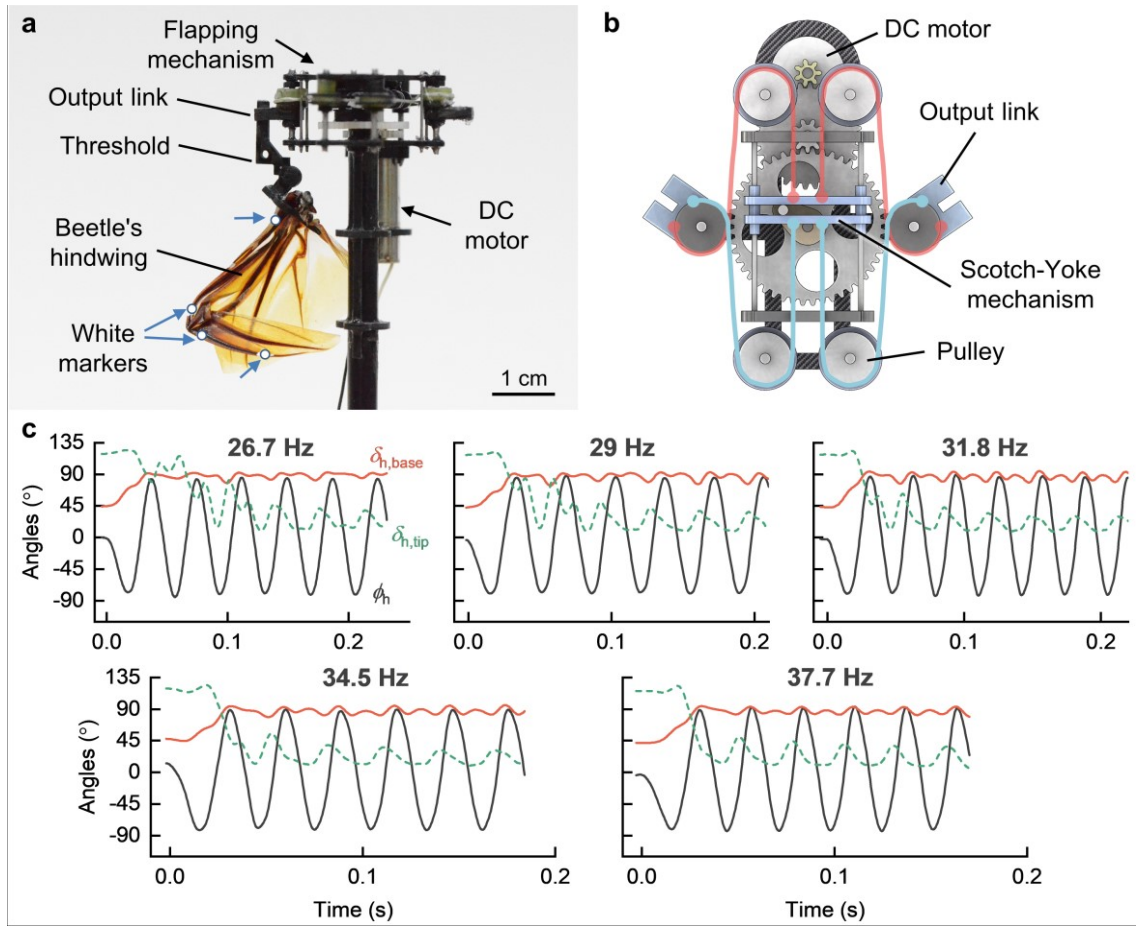

**Extended Data Fig. 2| Beetle's hindwing deployment experiments using a motor-driven flapping mechanism. a,** Experimental setup. **b**, Flapping mechanism design combining the Scotch-Yoke and pulley-string mechanisms to convert rotary motion of the DC motor to high-stroke flapping motion of the wing. **c**, Hindwing deployment kinematics at various flapping frequencies.

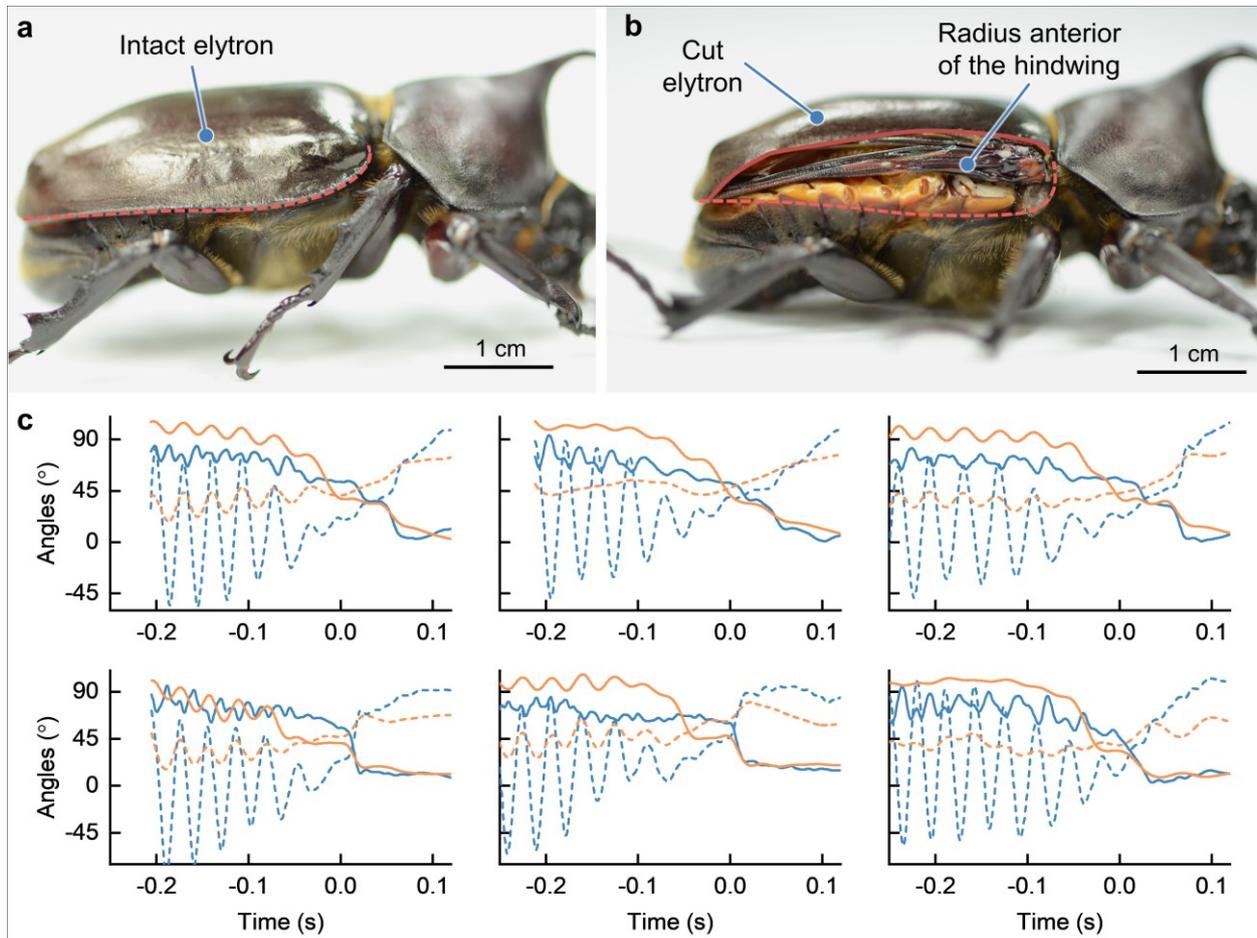

**Extended Data Fig. 3| Rhinoceros beetles use their elytra to depress the hindwings to the resting position after flight. a,b**, When at rest, the hindwings can be folded neatly inside the elytra. **c,** Stroke (dashed line) and elevation (solid line) angles of the elytron (orange) and hindwing (blue) during retraction. The time instant is set to 0 s when the elytron touches the hindwing while closing.

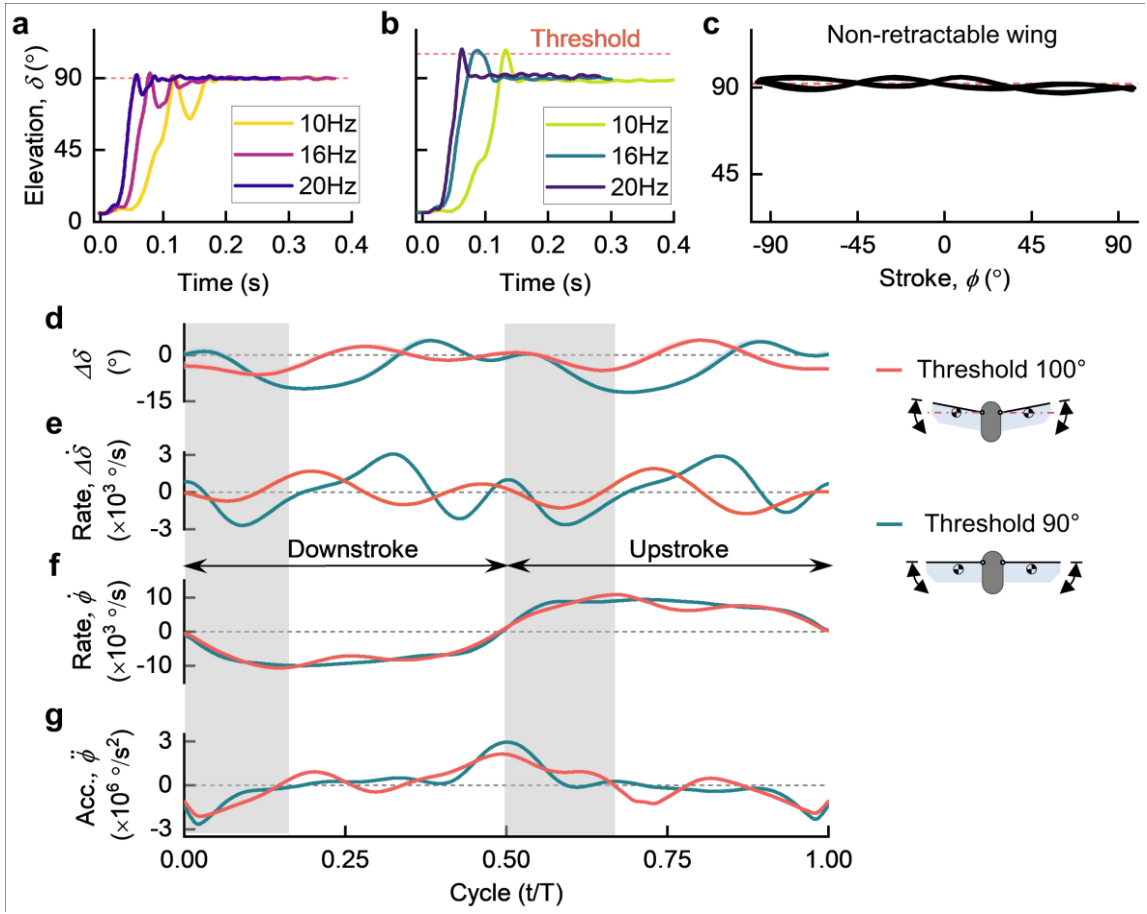

**Extended Data Fig. 4| Deployment kinematics of the robotic wings. a,b**, Without wing membrane, the centrifugal force keeps the wing on the plane normal to the flapping axis ($\delta = 90°$), even with the elevation threshold of $\delta_{threshold} = 110°$ (**b**). **c**, Wing tip trajectory of the non-retractable wing. The oscillation of the elevation angle is due to the bending of the leading-edge spar during flapping motion. **d,e**, Deviation of the elevation angles (**d**) from the threshold angles of 90° (cyan) and 100° (red) ($\Delta\delta = \delta - \delta_{threshold}$), and deviation rate (**e**). **f,g**, Stroke angular velocity (**f**) and acceleration (**g**). The downward movement of the wing is developed at the beginning of each stroke (the first half stroke denoted by the shaded area in d–g where the wing accelerates, similar to what was observed in the wing-inertia-only case.

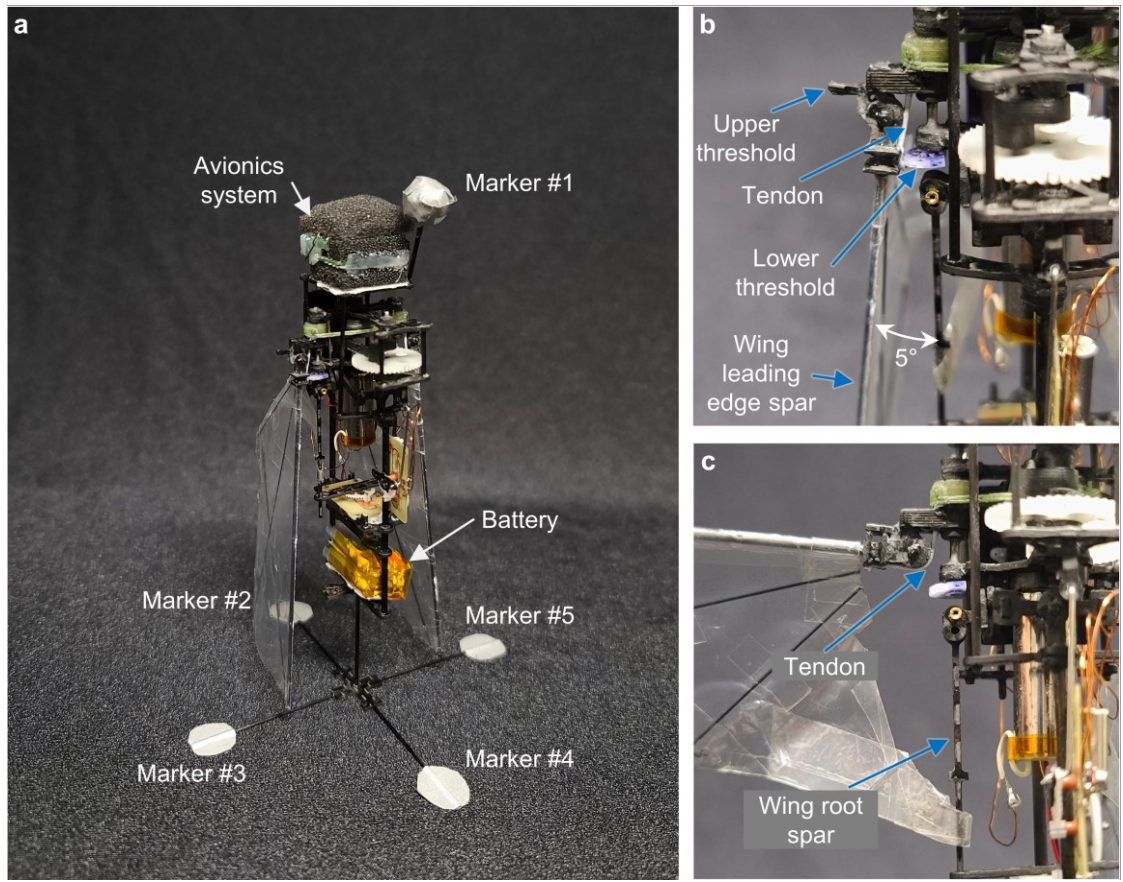

**Extended Data Fig. 5| The 18.2 gram flapping-wing robot with tracking markers used in flight experiments. a**, The markers added only 0.2 gram to the robot. The avionics system is covered with damping foam to reduce vibration noise during flapping flight and for protection. **b,c**, Close-up of the wing in folded (**b**) and extended (**c**) configurations. The wings remain folded at the lower threshold (about 5° from the wing root spar).

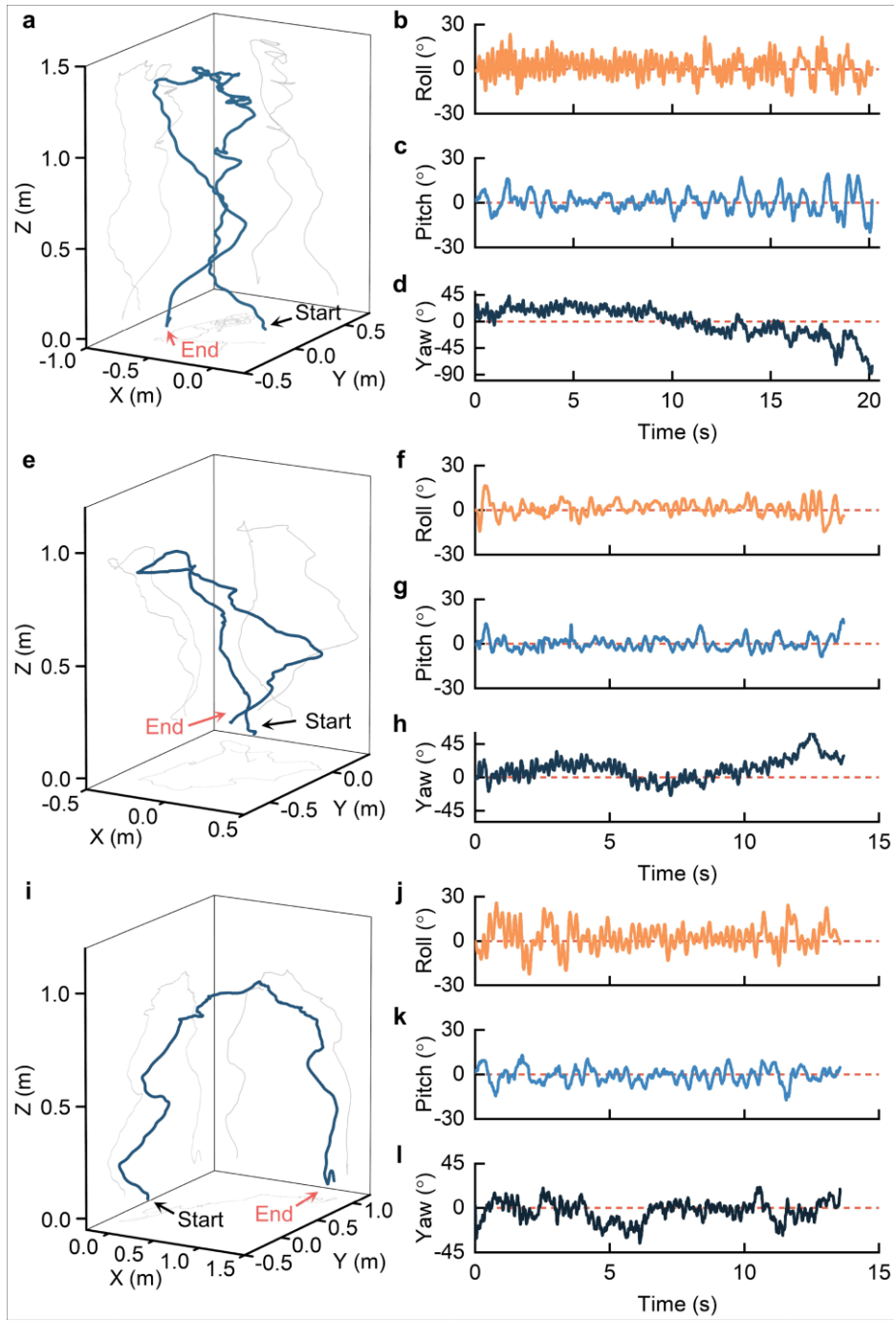

**Extended Data Fig. 6| Additional untethered flight experiments. a-d**, Test #1: flight trajectory (**a**), and roll (**b**), pitch (**c**), and yaw (**d**) attitude angles. **e-h**, Test #2: flight trajectory (**e**), and roll (**f**), pitch (**g**), and yaw (**h**) attitude angles. **i-l**, Test #3: flight trajectory (**i**), and roll (**j**), pitch (**k**), and yaw (**l**) attitude angles. Red dashed line denotes the reference.